\documentclass[twocolumn,epj]{svjour}
\usepackage{amsfonts}
\usepackage{amsmath}
\usepackage{amssymb}
\usepackage{graphicx}

\graphicspath{{./}{Figuras/}}

\begin{document}

\title{Gravitational collapse of a magnetized fermion gas with finite temperature}

\author{I. Delgado Gaspar\inst{1} \and A. P\'erez Mart\'\i nez\inst{2} \and Roberto A. Sussman\inst{3} \and A. Ulacia Rey\inst{3,4}}

\institute{
Instituto de Geof\'isica y Astronom\'ia (IGA), Calle 212 No 2906, La Lisa, La Habana, 11600, Cuba, \email{idelgado@iga.cu}
\and
Instituto de Cibern\'{e}tica, Matem\'{a}tica y F\'{\i}sica (ICIMAF), Calle E esq 15 No. 309 Vedado, La Habana, 10400, Cuba, \email{aurora@icmf.inf.cu}
\and
Instituto de Ciencias Nucleares, Universidad Nacional Aut\'onoma de M\'exico (ICN-UNAM). A. P. 70--543, 04510 M\'exico D. F., \email{sussman@nucleares.unam.mx}
\and
Instituto de Cibern\'{e}tica, Matem\'{a}tica y F\'{\i}sica (ICIMAF), Calle E esq 15 No. 309 Vedado, La Habana, 10400, Cuba, \email{alain@icmf.inf.cu}
}

\date{Received: date / Revised version: date}

\abstract{We examine the dynamics of a self--gravitating magnetized electron gas at finite
 temperature near the collapsing singularity of a Bianchi-I spacetime. Considering a general and appropriate and physically motivated initial conditions, we transform Einstein--Maxwell field
equations into a complete and self--consistent dynamical system amenable for numerical work. The resulting numerical solutions reveal the gas collapsing into both, isotropic (``point--like'')
and anisotropic (``cigar--like'') singularities, depending on the initial intensity of the magnetic field. We provide a thorough study of the near collapse behavior and interplay of all relevant state and kinematic variables: temperature, expansion scalar, shear scalar, magnetic field, magnetization and energy density. A significant qualitative difference in the behavior of the gas emerges in the temperature range $\hbox{T}\sim10^{4}\hbox{K}$ and  $\hbox{T}\sim 10^{7}\hbox{K}$.}


\maketitle

\section{Introduction}

Astrophysical systems provide an ideal scenario to examine the effects of strong magnetic fields
associated with self--gravitating sources under critical conditions. In such conditions, we expect non--trivial coupling between
gravitation and other fundamental interactions (strong, weak and electromagnetic), and from this interplay important
clues of their unification could emerge.

The presence and effects of strong magnetic fields in compact
objects (neutron, hybrids and quark stars) have been studied  in the
literature (see ~\cite{shapiro,EOS1991,EOS1996,EOS1999,Cardall:2000bs,EOS2002,EOS2007,Paulucci:2010uj},
and references quoted therein), assuming various types of equations of state (EOS) has been obtained
and considering in some of these papers numerical solutions of the equilibrium Tolman--Oppenheimer--Volkov (TOV) equation.

As proven in previous work~\cite{shapiro,Paulucci:2010uj,Martinez:2003dz,Felipe:2008cm,Ferrer:2010wz},
the presence of a magnetic field is incompatible with spherical symmetry and necessarily introduces
anisotropic pressures. However, by assuming that anisotropies and deviations from spherical symmetry
remain small, several authors~\cite{Paulucci:2010uj,Felipe:2008cm,Felipe:2010vr} have managed to compute
observable quantities of idealized static and spherical compact objects under the presence of strong
magnetic fields by means of TOV equations that incorporate these anisotropic pressures. Moreover, it is
evident that much less idealized models would result by considering the magnetic field and its associated
pressure anisotropies in the context of TOV equations under axial symmetric (or at least cylindrically symmetric)
geometries~\cite{Paulucci:2010uj}.

As an alternative (though still idealized) approach, and bearing in mind the relation between magnetic fields
and pressure anisotropy, we have examined the dynamics of magnetized self--gravitating Fermi gases as sources
of a Bianchi I space-time~\cite{Alain_e-,Alain_2,Alain_n}, as this is the simplest non--stationary geometry that
is fully compatible with the anisotropy produced by magnetic field source.

Evidently, a Bianchi I model is a completely inadequate metric for any sort of a compact object, as all geometric
and physical variables depend only on time (and thus it cannot incorporate any coupling of gravity with spatial
gradients of these variables). However, the use of this idealized geometry could still be useful to examine
qualitative features of the local behavior of the magnetized gas under special and approximated conditions.
Specifically, we aim at providing qualitative results that could yield a better understanding of the conditions
approximately prevailing near the center and the rotation axis of less idealized configurations, where the angular
momentum of the vorticity and the spatial gradients of the 4--acceleration and other key variables play a minor dynamical role.

The main objective of the present paper is to study the dynamical
evolution of a self-gravitating magnetized Fermi gas at finite
temperature, under the extremely critical conditions near a collapsing singularity. In particular, we examine the possibility that a finite temperature may produce the dynamical effect of ``slowing down'' or reversing the collapse of the magnetized gas. Also, we aim at verifying how a finite temperature affects the evolution of other thermodynamic variables, such as the magnetization and energy density. This work constitutes a continuation and enhancement of previous work~\cite{Alain_e-}.

The paper is organized as follow. In section II we derive the EOS for a dense magnetized electron gas at finite temperature.
In section III we lay out the dynamical equations for the evolution of our model by writing up the Einstein--Maxwell
system of equation for the specific source under consideration. In section
IV we introduce the local kinematic
variables, which are later used in section V to write the Einstein--Maxwell equations
as a system of non--linear autonomous differential equations, rewriting it afterwards in terms of physically motivated
dimensionless variables. The numerical analysis of the collapsing solutions and the discussion of the physical results
are given in the section  VI. Our conclusions are presented in sections VII.

\section{Magnetized Fermi gas as a source of a Bianchi I background geometry}

Homogeneous but anisotropic Bianchi I models are described by the Kasner metric\footnote{Unless specified otherwise, we use natural units.}

\begin{equation}\label{KasnerMetric}
ds^{2}=-dt^{2}+Q_{1}\left(t\right)^{2}dx^{2}+
Q_{2}\left(t\right)^{2}dy^{2}+Q_{3}\left(t\right)^{2}dz^{2},
\end{equation}
so that spatial curvature vanishes and all quantities depend only on time.
Assuming a comoving frame with coordinates
$x^a=\left[t,x,y,z\right]$ and 4-velocity $u^{a}=\delta_{t}^{a}$, the
energy--momentum tensor for a self--gravitating magnetized gas of free electrons is given by:
\begin{equation}
T^{\,\,a}_{b}=\left(U+P\right)u^{a}u_{b}+P\delta^{\,\,a}_{b}+\Pi^{\,\,a}_{b},
\quad P=p-\frac{2BM}{3},\label{eq:Tabtensorial}
\end{equation}
where $B$ is the magnetic field (pointing in the $z$ direction), $U$
is the energy density (including the rest energy of the electrons),
$M=-\left(\partial\Omega / \partial B \right)$
with $\Omega=\Omega_{e}+B^{2}/2$ and $\Omega_{e}$ is the
thermodynamical
potential of the electron gas, $P$ is the isotropic pressure
and $\Pi^{\,\,a}_{b}$ is the traceless anisotropic pressure tensor:
\begin{equation}\label{eq:Pi_aniso}
\Pi^{\,\,a}_{b}= diag \left[0,\Pi,\Pi,-2\Pi\right],\quad \Pi=-\frac{BM}{3},
\end{equation}
We can write the energy--momentum (\ref{eq:Tabtensorial}) tensor as:
\begin{equation}
T^{\,\,a}_{b}= \hbox{diag}\left[-U,P_{\bot},P_{\bot},P_{\|}\right],\label{eq:TabMatricial}
\end{equation}
which respectively identifies $P_{\bot}$ and $P_{\|}$ as
the pressure components perpendicular and parallel to the magnetic field.
Notice that the anisotropy in $T^{\,\,a}_{b}$ is produced by the magnetic
field $B$. If this field vanishes, the energy--momentum tensor
reduces to that of a perfect fluid with isotropic
pressure (an ideal gas of electrons complying with Fermi--Dirac statistics).

The EOS for this magnetized electron gas can be given
in the following form \cite{EOS1999,Ferrer:2010wz,Can1}:
%
\begin{eqnarray}
P_{\|}&=& -\Omega_{e}-BM_{e}+\frac{B^{2}}{2},\label{eq:Pzz}
\\
P_{\bot}&=&-\Omega_{e}-\frac{B^{2}}{2},\label{eq:Pxx}
\\
U&=&-\Omega_{e}+TS+\mu N+\frac{B^{2}}{2}.
\end{eqnarray}
%
where $M_{e}=-\left(\partial\Omega_{e} / \partial B \right)$ is
the electron-system magnetization,
$S=-\left(\partial\Omega_{e} / \partial T \right)$ is the entropy,
$N=-\left(\partial\Omega_{e} / \partial \mu \right)$ is the
particle number density and the term $B^{2}/2$ is Maxwell's classical
magnetic contribution. We note that equations (\ref{eq:Pzz}) and (\ref{eq:Pxx}) can be re--arranged as \cite{EOS1999,Martinez:2003dz}:

\begin{equation}
P_{\bot}=P_{\|}-B\emph{M},\label{eq:Pxx-Pzz}
\end{equation}

The thermodynamic potential $\Omega_{e}$ can be written as the summ:
\begin{equation}\label{contrib Omega}
    \Omega_{e}=\Omega_e^{QFT}+\Omega_e^{SQFT},
\end{equation}
where we identify the vacuum term $\Omega_e^{QFT}$, independent of the temperature and chemical potential, and the statistical term,
$\Omega_e^{SQFT}$. As is known,
$\Omega_e^{QFT}$ has non-field-dependent ultraviolet divergencies, hence
after renormalizing the Schwinger expression we obtain \cite{Schwinger}:

\begin{eqnarray}
\nonumber
\Omega_e^{QFT}(B)=-\frac{1}{8\pi^2}\int_{0}^\infty \frac{ds}{s^{3}}
\exp\left(-m_e^{2}s\right)\times
\\
\times\left(esB \coth\left(esB\right)
-1-\frac{\left(esB\right)^{2}}{3}\right),  \label{Omega V}
\end{eqnarray}
\begin{eqnarray}
\nonumber
\Omega_e^{SQFT}(B,T,\mu)&=&
\\
-\frac{eB}{4\pi^2\beta}\int_{-\infty}^\infty
dp_3&&
\!\!\!\!\!
\sum_{l=0}^{\infty}d(l)\ln
\left(1+e^{-\beta(\epsilon_l-\mu)}\right), \label{Omega Statitical}
\end{eqnarray}
where $d(l)=2-\delta_{l0}$ is the spin degeneracy
of Landau levels with $l\neq0$, $m_e$ is the electron mass,
$\epsilon_l=p_3^{2}+2|eB|l+m_e^{2}$ and
$\beta=1/T$.

Finally, we remark that in a high density regime (which we shall consider henceforth) the
contributions to the EOS of the vacuum term and the
Maxwell term ({\it i.e.} $\Omega_e^{QFT}$ and $B^{2}/2$) can be
neglected whenever the magnetic field is less than critical
 ({\it i.e.} $B<B_{c}=m_e^{2}/e$). Thus, in subsequent calculations the
leading contribution to the EOS will come from $\Omega_e^{SQFT}$
(see Appendix A).

\section{Einstein--Maxwell equations}

Since we are interested in the critical relativistic regimen, the
dynamics of the magnetized gas whose EOS we have described in the
previous section must be studied through the Einstein field
equations in the framework of General Relativity:
\begin{equation}
G_{\mu\nu}=R_{\mu\nu}-\frac{1}{2}Rg_{\mu\nu}=\kappa
T_{\mu\nu},\label{eq:Ecuac Einstein}
\end{equation}
together with the balance equations of the energy-momentum
tensor and Maxwell's equations,
\begin{eqnarray}
T^{\mu\nu}\,_{;\nu}=0,\label{eq:conseracion E-P}
\\
F^{\mu\nu}\,_{;\nu}=0,  \qquad F_{\left[\mu\nu;\alpha\right]}=0,
\label{eq:ecuac d Maxwell}
\end{eqnarray}
where $\kappa=8\pi G_{N}$ and $G_{N}$ is Newton's gravitational
constant, while square brackets denote anti-symmetrization in (\ref{eq:ecuac d Maxwell}).

Assuming absence of annihilation/creation processes, so that
particle numbers are conserved, leads to the following
conservation equation:
\begin{equation}
n^{\alpha}\,_{;\alpha}=0,\qquad n^{\alpha}=N \,u^{\alpha},\label{eq:ecuac conservacion No. d particulas}
\end{equation}
where $N$ es the particle number density. From
the field equations (\ref{eq:Ecuac Einstein}) we obtain:
\begin{eqnarray}
-G^{\,\,x}_{x}=\frac{\dot{Q_{2}}\dot{Q_{3}}}{Q_{2}Q_{3}}+\frac{\ddot{Q_{2}}}{Q_{2}}+\frac{\ddot{Q_{3}}}{Q_{3}}=-\kappa P_{\bot},\label{eq:Gxx}
\\
-G^{\,\,y}_{y}=\frac{\dot{Q_{1}}\dot{Q_{3}}}{Q_{1}Q_{3}}+\frac{\ddot{Q_{1}}}{Q_{1}}+\frac{\ddot{Q_{3}}}{Q_{3}}=-\kappa P_{\bot},\label{eq:Gyy}
\\
-G^{\,\,z}_{z}=\frac{\dot{Q_{1}}\dot{Q_{2}}}{Q_{1}Q_{2}}+\frac{\ddot{Q_{1}}}{Q_{1}}+\frac{\ddot{Q_{2}}}{Q_{2}}=-\kappa P_{\|},\label{eq:Gzz}
\\
-G^{\,\,t}_{t}=\frac{\dot{Q_{1}}\dot{Q_{2}}}{Q_{1}Q_{2}}+\frac{\dot{Q_{1}}\dot{Q_{3}}}{Q_{1}Q_{3}}+
\frac{\dot{Q_{2}}\dot{Q_{3}}}{Q_{2}Q_{3}}=\kappa U.\label{eq:Gtt}
\end{eqnarray}
where $\dot{Q}=Q_{;\alpha}u^{\alpha}=Q_{,t}$. From the conservation
of the energy--momentum tensor (\ref{eq:conseracion E-P}) we obtain:
\begin{equation}
\dot{U}=-\left(\frac{\dot{Q_{1}}}{Q_{1}}+\frac{\dot{Q_{2}}}{Q_{2}}\right)\left(P_{\bot}+U\right)-
\frac{\dot{Q_{3}}}{Q_{3}}\left(P_{\|}+U\right).
\end{equation}
Maxwell's equations (\ref{eq:ecuac d Maxwell}) yield:
\begin{equation}
\frac{\dot{Q_{1}}}{Q_{1}}+\frac{\dot{Q_{2}}}{Q_{2}}+\frac{1}{2}\frac{\dot{B}}{B}=0,
\end{equation}
and from the particle number conservation
(\ref{eq:ecuac conservacion No. d particulas}) leads to:
\begin{equation}
\dot{N}+\left(\frac{\dot{Q_{1}}}{Q_{1}}+\frac{\dot{Q_{2}}}{Q_{2}}+\frac{\dot{Q_{3}}}{Q_{3}}\right)N=0.\label{eq:eta}
\end{equation}

\section{Local kinematic variables}

Einstein--Maxwell field equations are second order system of
ordinary differential equations (ODE's). In order to work with
a first order system of ODE's, it is useful and convenient to
rewrite these equations in terms of covariant kinematic variables
that convey the geometric effects on the kinematics of local
fluid elements through the covariant derivatives of $u^\alpha$.
For a Kasner metric in the comoving frame endowed with a
normal geodesic 4--velocity, the only non--vanishing
kinematic parameters are the expansion scalar, $\Theta$,
and the shear tensor $\sigma_{\alpha\beta}$:
\begin{equation} \Theta=u^{\alpha}\,_{;\alpha}\,,\quad \sigma_{\alpha\beta}=u_{(\alpha;\beta)}-\frac{\Theta}{3}h_{\alpha\beta}\,,\end{equation}
where $h_{\alpha\beta}=u_{\alpha}u_{\beta}+g_{\alpha\beta}$ is
the projection tensor and rounded brackets denote symmetrization.
These parameters take the form:
\begin{equation}
\Theta=\frac{\dot{Q_{1}}}{Q_{1}}+\frac{\dot{Q_{2}}}{Q_{2}}+\frac{\dot{Q_{3}}}{Q_{3}}\,,\label{eq:def theta}
\end{equation}
\begin{equation}
\sigma^{\,\,\alpha}_{\beta}=\hbox{diag}\,\left[\sigma^{\,\,x}_{x},\sigma^{\,\,y}_{y},\sigma^{\,\,z}_{z},0\right]
=\hbox{diag}\,\left[\Sigma_{1},\Sigma_{2},\Sigma_{3},0\right],\label{eq:def sigma}
\end{equation}
where:
\begin{eqnarray}
\nonumber
\Sigma_{a}=\frac{2}{3}\frac{\dot{Q_{a}}}{Q_{a}}-\frac{1}{3}\frac{\dot{Q_{b}}}{Q_{b}}&-&
\frac{1}{3}\frac{\dot{Q_{c}}}{Q_{c}},
 \\
&{}&\!\! a\neq b\neq c\,\left(a,b,c=1,2,3\right).
\label{eq: componentes}
\end{eqnarray}
The geometric interpretation of these parameters is
straightforward:  $\Theta$
represents the isotropic rate of change of the 3--volume of
a fluid element, while
$\sigma^{\,\,\alpha}_{\beta}$ describes its rate of local
deformation along different spatial directions given
by its eigenvectors.
Since the shear tensor is traceless:
$\sigma^{\,\,\alpha}_{\alpha}=0$,
it is always possible to eliminate any one of the three
quantities $\left(\Sigma_{1},\Sigma_{2},\Sigma_{3}\right)$ in
terms of the other two. We choose to eliminate $\Sigma_{1}$
as a function of $\left(\Sigma_{2},\Sigma_{3}\right)$.

\section{Dynamical equations}

By using equations (\ref{eq:def theta}) and (\ref{eq: componentes})
we can re--write the second derivatives of the metric functions
in (\ref{eq:Gxx}), (\ref{eq:Gyy}) y (\ref{eq:Gzz}) as first order
derivatives of $\Theta$, $\Sigma_{2}$ and $\Sigma_{3}$.
After some algebraic manipulations it is possible to
transform equations (\ref{eq:Gxx})-(\ref{eq:eta}) as a first
order system of autonomous ODE's:
\begin{eqnarray}
\dot{\Sigma}_{2}&=&\frac{\varkappa}{3}\left(P_{\bot}-P_{\|}\right)-\Theta\Sigma_{2},\label{eq:sistema1}
\\
\dot{\Sigma}_{3}&=&\frac{2\varkappa}{3}\left(P_{\|}-P_{\bot}\right)-\Theta\Sigma_{3},\label{eq:sistema2}
\\
\dot{B}&=&2B\left(\Sigma_{3}-\frac{2}{3}\Theta\right),\label{eq:sistema3}
\\
\dot{\mu}&=&\frac{1}{\hbox{Det}}\left[ f_{1}N_{,T}-f_{2}U_{,T}\right],\label{eq:sistema4}
\\
\dot{T}&=&\frac{1}{\hbox{Det}}\left[ f_{2}U_{,\mu}-f_{1}N_{,\mu}\right],\label{eq:sistema5}
\end{eqnarray}
together with the following constraint:
\begin{equation}
\varkappa U=-\left(\Sigma_{2}\right)^{2}-\Sigma_{2}\Sigma_{3}-
\left(\Sigma_{3}\right)^{2}+\frac{\Theta^{2}}{3},\label{eq: vinculoSAdi}
\end{equation}
where:
\begin{eqnarray}
\hbox{Det} &\equiv& U_{,\mu}N_{,T}-U_{,T}N_{,\mu},
\\
 f_{1} &=& \left(\Sigma_{3}-\frac{2\Theta}{3}\right)\left(P_{\bot}+U\right)\nonumber
\\
 &{}&\quad-\left(\Sigma_{3}+\frac{\Theta}{3}\right)\left(P_{\|}+U\right)-U_{,B}\dot{B},
\\
 f_{2} &=& \Theta N+N_{,B}\dot{B}.
\end{eqnarray}

We introduce below the following dimensionless evolution parameter,
\begin{equation}\label{adim t,H,S2,S3}
    \frac{d}{d\tau}=\frac{1}{H_{0}}\frac{d}{dt},\quad
    \mathcal{H}=\frac{\Theta}{3H_{0}},\quad
    S_{2}=\frac{\Sigma_{2}}{H_{0}},\quad
    S_{3}=\frac{\Sigma_{3}}{H_{0}},
\end{equation}
\begin{equation} \label{adim b,mu,T}
   b=\frac{B}{B_{c}},\qquad
   \tilde\mu=\frac{\mu}{m_{e}},\qquad
   \phi=\frac{T}{m_{e}},
\end{equation}
and write the EOS in the form:
\begin{eqnarray}
  U &=& \lambda\,\Gamma_{U}\left(\beta,\tilde \mu,\phi\right), \\
  P_{\bot} &=& \lambda\,
\Gamma_{\bot}\left(\beta,\tilde \mu,\phi\right), \\
  P_{\|} &=& \lambda\,\Gamma_{\|}\left(\beta,\tilde \mu,\phi\right), \\
  N &=& \left(\lambda/m_{e}\right)\,\Gamma_{\eta}\left(\beta,\tilde \mu,\phi\right),
  \label{U adim}
\end{eqnarray}
where $\lambda=8\pi m_{e}/(\lambda_{c}^{3})$ with  $\lambda_{c}$ the Compton wavelength of the electron
and $H_{0}$ is a constant with inverse length units that sets the
characteristic length scale of the system, which we have chosen as
$3H_{0}^{2}=\kappa\lambda\Rightarrow H_{0}=0.86 \times 10^{-10}\,\hbox{m}^{-1}$, so
that $1/H_{0}\cong 1.15\times 10^{10}\, \hbox{m}$ is of the order
of magnitude of
an astronomic unit. It indicates that our simplified model is
examined on local scales smaller than cosmic scales.
In cosmological sources and models \cite{MTWGravitation}
$H_{0}=0.59 \times 10^{-26}\,\hbox{m}^{-1}$ would play
the role of the Hubble
scale constant, this value is a much greater length scale. The
functions $S_{2}$ and $S_{3}$ are the components of the shear tensor
normalized with this scale, while $\tau$ is the dimensionless time.

Substituting (\ref{adim t,H,S2,S3})--(\ref{U adim}) into the system (\ref{eq:sistema1})--(\ref{eq:sistema5})
we obtain:
\begin{eqnarray}
S_{2,\tau}&=&\Gamma_{\perp}-\Gamma_{\|}-3\mathcal{H}S_{2},\label{eq:sistadimen-a}
\\
S_{3,\tau}&=&2\left(\Gamma_{\|}-\Gamma_{\perp}\right)-3\mathcal{H}S_{3},
\\
b_{,\tau}&=&2b\left(S_{3}-2\mathcal{H}\right),
\\
\tilde{\mu}_{,\tau}&=&\frac{1}{\hbox{Det}}\left(\Gamma_{N,\phi}\widetilde{f}_{1}+\Gamma_{U,\phi}\widetilde{f}_{2}\right),
\\
\phi,\tau&=&-\frac{1}{\hbox{Det}}\left(\Gamma_{N,\tilde \mu}\widetilde{f}_{1}+\Gamma_{U,\tilde \mu}\widetilde{f}_{2}\right),\label{eq:sistadimen-e}
\end{eqnarray}
while the constraint (\ref{eq: vinculoSAdi}) becomes,
\begin{equation}
3\Gamma_{U} = -S{}_{2}^{2}-S{}_{3}^{2}-S_{2}S_{3}+3\mathcal{H}^{2},\label{eq:vincadimen}
\end{equation}
and the auxiliary parameters of the previous system take the form:
\begin{eqnarray}
\widetilde{f}_{1}&=&\left(S_{3}-2\mathcal{H}\right)\left(\Gamma_{\perp}-2\Gamma_{U,b}b\right)-
\left(S_{3}+\mathcal{H}\right)\Gamma_{\parallel}-3\Gamma_{U}\mathcal{H}, \nonumber
\\
& &\quad\;\widetilde{f}_{2}=3\mathcal{H}\Gamma_{N}+
2\Gamma_{N,b}\left(S_{3}-2\mathcal{H}\right),
\\
& &\quad\;\hbox{Det}=\Gamma_{U,\tilde \mu}\Gamma_{N,\phi}-
\Gamma_{U,\phi}\Gamma_{N,\tilde \mu}\,. \nonumber
\end{eqnarray}

These differential equations form a complete and self-consistent system
whose numeric integration fully determines the variables
$\mathcal{H},S_{2}$, $S_{3}$, $b$, $\tilde \mu$, $\phi$,
thus allowing us to study
the dynamical evolution of a local volume element
of a gas of magnetized electrons with finite temperature.
In particular, all thermodynamical magnitudes, such as the particle number density,
the magnetization, the energy density,
the entropy and the pressures, are functions of $b(\tau)$, $\tilde \mu(\tau)$, $\phi(\tau)$
and thus can be determined from the numerical solution of this system.
On the other hand, $\mathcal{H},S_{2}$, $S_{3}$ provide
all the necessary information
on the kinematic evolution of volume element, and in particular, its
proper volume and the metric functions.

Since $\mathit{V}=\sqrt{-\det g_{\alpha\beta}}=Q_{1}Q_{2}Q_{3}$,
we obtain by means of (\ref{eq:def theta}) and
(\ref{adim t,H,S2,S3}) the local volume in terms of ${\mathcal{H}}$:
\begin{equation}
\mathit{V}\left(\tau\right)=\mathit{V}\left(0\right)\exp\left(3\intop_{\tau=0}^{\tau}\mathcal{H}d\tau\right),\label{eq: V(tau)}
\end{equation}
where we remark that the sign of
$\mathcal{H\left(\tau\right)}$ implies expansion
if ${\mathcal{H\left(\tau\right)}}>0$, and collapse
if ${\mathcal{H\left(\tau\right)}}<0$. Besides
this point, equations (\ref{eq:def theta}) and
(\ref{eq: componentes}) lead to:
\begin{equation}
Q_{i}(\tau)=Q_{i}(0)\exp\left[\,\,\intop_{\tau=0}^{\tau}\left(\mathcal{H}+S_{i}\right)d\tau\right],\qquad\quad\left(i=1,2,3\right).\label{eq: Q(tau)}
\end{equation}
where $S_{1}=-\left(S_{2}+S_{3}\right)$. In the following section we undertake the numerical study of the
system (\ref{eq:sistadimen-a})-(\ref{eq:vincadimen}), focusing
specifically on the the collapsing regime.

\section{Numeric analysis and physical interpretation}

For the numerical study we assume a magnetized electron gas at high
density: $\tilde \mu_e(0)=2$, which
means that the chemical potential is $2m_{e}$. The initial values
of the magnetic field and
temperature were chosen in the ranges  $b(0)\sim10^{-5}$
to $b(0)\sim 10^{-4}$ and $\phi\sim10^{-6}$ to $\phi\sim 10^{-3}$
respectively
\footnote{
For instance $B\thicksim 10^{8}\hbox{G}$ ($b\thicksim 10^{-5}$),
$T\sim 10^{7}\hbox{K}$ ($\phi\thicksim 10^{-3}$) and
$\mu \simeq 1 \hbox{MeV}$ ($\tilde \mu=2$)
correspond to the physical parameters of a white dwarf type astrophysical object.}.
 Together with $\mathcal{H}\left(0\right)<0$, we consider the
 conditions $S_{2}(0)=0$, $S_{3}(0)=0,+1$, which
 correspond to the cases with zero initial deformation and initial
 deformation (shear) in the direction of $z$
 axis, respectively. The calculation has been done using the
 fourth-order Runge--Kutta method with the local
 truncation relative error less than $10^{-6}$.

\begin{figure}
\includegraphics[scale=0.35]{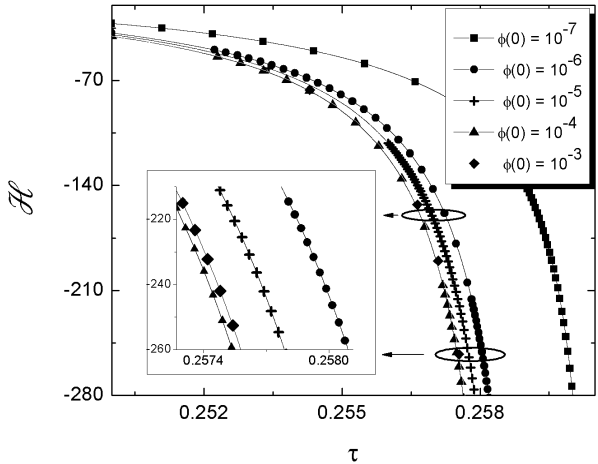}
\caption{\label{fig:G1} Numerical solutions for the expansion scalar $\mathcal{H}\left(\tau\right)$. The
 box in the lower corner is an amplification of the graphic.}
\end{figure}

The numerical solutions for the function $\mathcal{H}$ for the
assumed values $\mathcal{H}(0)$ show that
$\mathcal{H}\rightarrow-\infty$, which implies that the
volume element evolves to a singularity (see
equation (\ref{eq: V(tau)})). This is exemplified
in figure \ref{fig:G1}, where numerical
solutions are displayed for the expansion scalar.
These curves correspond to different values of the
initial temperature in the range $\phi(0)=10^{-7}$ to
$\phi(0)=10^{-3}$, fixing the rest of the initial
conditions on the values $\tilde \mu(0)=2$, $b(0)=5\times10^{-5}$,
$S_{2}(0)=0$ and $S_{3}(0)=1$. Notice that
in this regime (as given by these initial conditions) where the high
densities are dominant, we do not obtain
a direct relation between the values of initial temperatures and the
collapse time. However, the collapsing time diminishes when
the values of the initial magnetic field increases, which
agrees with the results obtained in \cite{Alain_e-}, where $T=0$
was assumed.

In all examined configurations the strength of the magnetic field
diverges during the collapse
of the volume element, an expected development due to compression of the
magnetic fluid.
Since the energy--momentum tensor (\ref{eq:TabMatricial}) takes
the perfect fluid form ($P_{||}=P_\perp$) for
zero magnetic field, we can identify the magnetic field as the
factor introducing anisotropy in the dynamical behavior
of the fermionic gas. In particular, this anisotropy in the
pressures ($P_{||}\ne P_\perp$) must yield different
evolution in different directions, which must be evident
in a critical stage such as the collapsing regime.
When the evolution occurs in such a way that all
metric coefficients tend to zero, the singularity is necessarily isotropic
``point--like'', whereas when two  metric coefficients  evolve
to zero and the third one to $+\infty$ the result is an anisotropic singularity,
``cigar-like'' (see the definitions of these types of singularities in
\cite{dtipo_d_singul}). It follows from (\ref{eq: Q(tau)}) that
the evolution of the terms $S_{i}+\mathcal{H}$ to $\pm\infty$
implies that the metric
coefficient  $Q_{i}$ evolves to either $+\infty$ or $0$, which
determines the
type of singularity. Intuitively, we expect an isotropic point--like
singularity if the pressure is isotropic, as pressure diverges in
all directions, but a large pressure anisotropy (which necessarily
corresponds to large magnetic field) should
lead to a qualitatively different direction dependent critical
behavior of the pressure, which should result in an
anisotropic cigar--like singularity characterized by the divergence
of only the pressure parallel to the magnetic
field.

\begin{figure}
\includegraphics[scale=0.35]{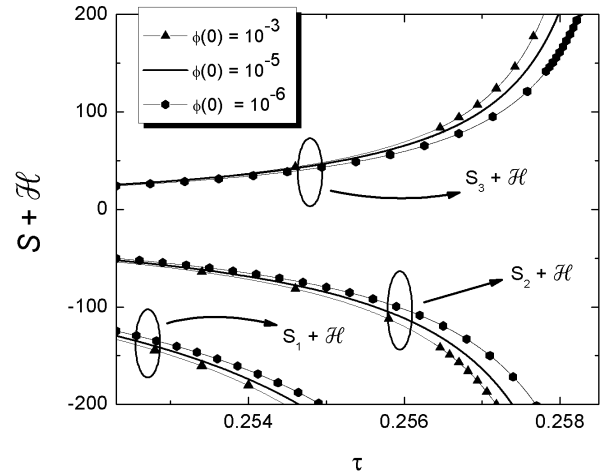}
\caption{\label{fig S+H} The plots of the functions $S_{i}+\mathcal{H}$ for $i=1,2,3$. The
system has an initial deformation (shear) in the same direction of the
magnetic field: $z$ axis. The collapse is in the form of
a ``cigar--like'' singularity in the $z$ direction.}
\end{figure}
\begin{figure}
\includegraphics[scale=0.35]{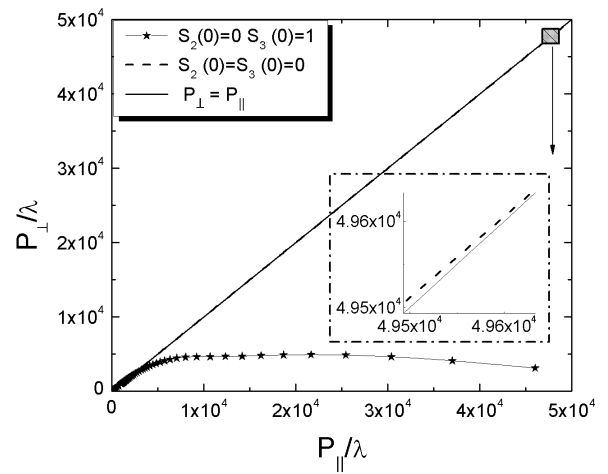}
\caption{\label{fig presiones} Pressure parallel and perpendicular to the
magnetic field. We plot $P_{||}/\lambda$ vs $P_{\perp}/\lambda$ for
initial shear  $S_{2}(0)=S_{3}(0)=0$ and $S_{2}(0)=0 \wedge S_{3}(0)=1$ taking $\phi\left(0\right)=10^{-3}$.
$\lambda=8\pi m_{e}/(\lambda_{c}^{3})$. The case $b=0$ is
shown for reference.}
\end{figure}

In the numerical calculations whose initial
conditions assume zero shear: $S_{1}(0)=S_{2}(0)=S_{3}(0)=0$, we always
obtain a point--like singularity, independently of the selected
values for the remaining
initial conditions. However, when the initial deformation (shear)
is positive in the direction of the magnetic field ($z$ axis), we
obtain a cigar--like singularity.
Figure \ref{fig S+H} displays the solutions of
$S+\mathcal{H}$ for initial
conditions: $\tilde \mu(0)=2$, $b(0)=5\times10^{-5}$,
$S_{3}\left(0\right)=1$, $S_{2}\left(0\right)=0$ and
$\phi\left(0\right)=10^{-6},\,10^{-6},\,10^{-3}$. The nine curves
displayed in the figure correspond to three sets
of functions: $S_{i}+\mathcal{H}\;\;\left(i=1,2,3\right)$, each one
corresponding to three different values of
the temperature: $\phi\left(0\right)=10^{-6},\,10^{-5},\,10^{-3}$.
As we mentioned previously, in these cases the gas collapses into
a cigar--like
singularity in the direction of the magnetic field, and when the initial shear is zero the three curves $S_{i}+\mathcal{H}$ tend to $-\infty$ for each initial value of temperature.

In figure \ref{fig presiones} we compare the behavior of
the anisotropic pressure in the directions parallel vs perpendicular during the collapse.
The dashed line corresponds to the case
with zero initial shear, while the star--gray line
 corresponds to the case of initial positive shear in the $z$ direction.
 In both cases the remaining initial condition were fixed as:
 $\phi\left(0\right)=10^{-3}$,
$\tilde\mu(0)=2$ and $b(0)=5\times10^{-5}$.
As shown by the figure, the pressures exhibit
the expected behavior: for initial zero shear we have
$P_{||}\approx P_\perp$ and a point singularity, but
for initial positive shear in the $z$ direction
we have a highly anisotropic
evolution $P_{||}\gg P_\perp$,
which signals a cigar--like singularity. The case $P_{||}= P_\perp$
with zero magnetic field is shown for
comparison (solid line).

For initial values in the range $\phi(0)\sim10^{-6}-10^{-4}$ the temperature decreases
as the collapse proceeds, while for $\phi(0)\sim10^{-3}$
and higher values, it increases.
This behavior of temperature can be
explained by considering that for small initial values
($\phi(0)\sim10^{-6}-10^{-4}$) the effects of the magnetic
field (alignment of electron spins
with the magnetic field)
predominate over the effects of temperature
(random motion) throughout the whole
evolution. Therefore, we argue that the system
``cools down'' (temperature decreases) during the collapse
process as the
magnetic field increases. On the other hand,
for high initial temperatures ($\phi\left(0\right)=10^{-3}$), the
system is ``heated'' (temperature increases) during the collapse,
which shows
the predominance of the thermal effect of
temperature over the magnetic field effect.

\begin{figure}
\includegraphics[scale=0.35]{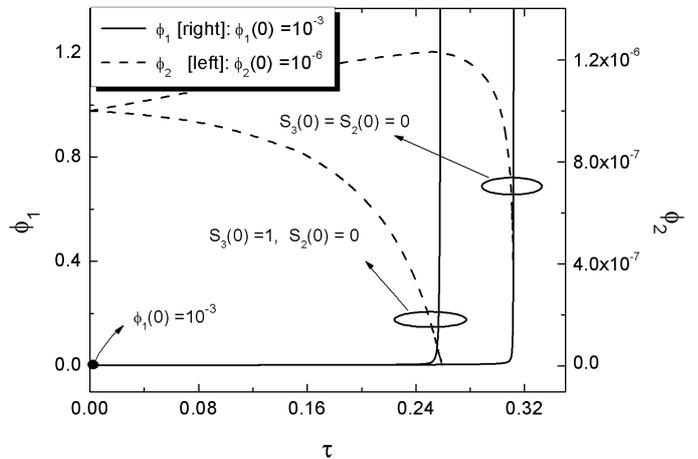}
\caption{\label{fig temperatura} Temperature of the system. The left hand side of the graph displays the
 temperature values ($\phi_{1}$) corresponding to the two solid curves. The right hand side displays the
 temperature values ($\phi_{2}$) corresponding to the two dashed curves.}
\end{figure}

We display in figure \ref{fig temperatura} two dashed
curves for the evolution of the temperature from
the initial value $\phi\left(0\right)=10^{-3}$, together
with solid curves that depict temperatures
starting at $\phi\left(0\right)=10^{-6}$.
In both cases the initial conditions correspond to zero
initial shear and positive deformation in the $z$ axis.
The initial values of magnetic field and chemical
potential were fixed at $b(0)=5 \times 10^{-5}$ and $\tilde \mu(0)=2$
respectively. Notice that the symmetric configuration
delays the collapse (yields a longer collapsing time), which may
indicate a connection with stability of local fluid
elements in compact objects. This could provide an important clue on
the stability of a compact object made of a dense magnetized gas.
However, verifying this possibility is beyond the
scope of this article.

\begin{figure}
\includegraphics[scale=0.35]{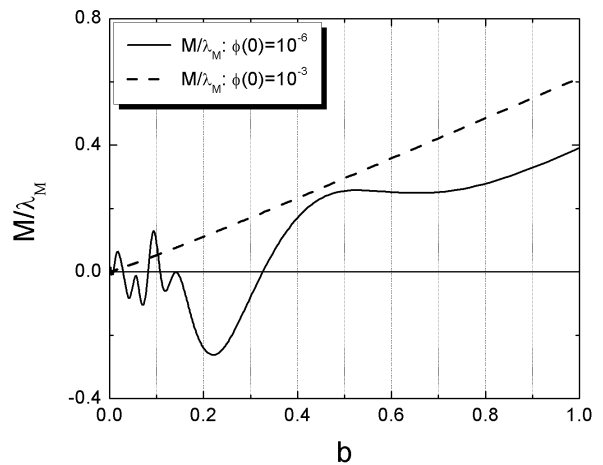}
 \caption{\label{fig M} Magnetization  versus magnetic field. The  dashed curve (the solid curve) represents the magnetization taking
 $\phi\left(0\right)=10^{-3}$ ($\phi\left(0\right)=10^{-6}$).}
\end{figure}

\begin{figure}
\includegraphics[scale=0.35]{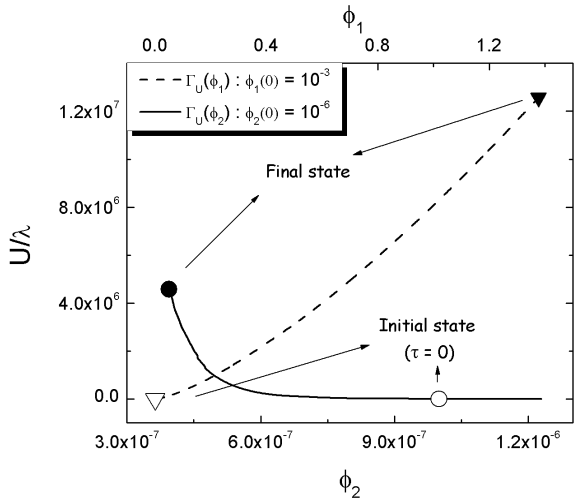}
 \caption{\label{fig U}Energy density versus temperature. The temperature values of the curve
draws with solid line should be observed in the lower axis ($\phi_{2}$) and
in the top axis ($\phi_{1}$), those of the curve drawn with dashed line.}
\end{figure}

We examine now the behavior of the magnetization and the
energy density of the system during the
collapse. We remark that the $\tau$--depending functions:
$b,\; \tilde \mu,\; \phi$ follow from the numerical solution
of the system
(\ref{eq:sistadimen-a})--(\ref{eq:sistadimen-e}) and
(\ref{eq:vincadimen}), which allows us to compute the magnetization
$M\left(b, \;\tilde \mu, \;\phi\right)$ and $U\left(b, \;\tilde \mu,
\;\phi\right)$ for all values of $\tau$. The figures
figures \ref{fig M} and \ref{fig U} have been done
with the initial conditions: $b\left(0\right)=5\times10^{-5}$,
$S_{2}\left(0\right)=S_{3}\left(0\right)=0$.

Figure \ref{fig M} depicts the behavior of the magnetization
vs the magnetic field.  Notice how the gas exhibits the de Haas-van
Alphen oscillations for small values of
the magnetic field and temperature. These oscillations
have been observed in many magnetized systems, and
are a direct consequence of the
Pauli exclusion principle and the discreteness of the
spectrum \cite{Haas-van Alphen}.
When the gas starts from initial values $\phi\left(0\right)=10^{-6}$ the magnetization has an oscillatory behavior, with
these oscillations disappearing as the magnetic field increases.
As a contrast, for initial values $\phi\left(0\right)=10^{-3}$,
the system is
heated as the collapse proceeds and the magnetization
has a non--oscillatory behavior.
These results, which were obtained in a time-dependent system,  agree
with those obtained in \cite{TAndersen,Oliver},
where the authors also studied the magnetization of an
electron gas at finite temperature. This agreement shows how the time evolution that we have studied preserves the properties of the same system when it is time--independent.

In all the studied cases the energy density increases as
the collapse proceeds, though this fact does not imply
that temperature increases. As we have seen above
for $\phi\left(0\right)=10^{-3}$, the system is heated
during the collapse, which
yields a monotonic behavior of the energy density
with respect to the temperature. On the other hand, for
$\phi\left(0\right)=10^{-6}-10^{-4}$ the system evolves
in such way that temperature decreases in time as energy
density increases, without the observed monotonic behavior of the
energy density vs the temperature. In this latter case the chemical potential and the
particle density increase, even if the
temperature decreases and the magnetic field increases, Hence, the energy density
also grows during the collapse.
Figure \ref{fig U} displays two curves of the energy density vs temperature, with the dashed and solid line respectively  corresponding to $\phi\left(0\right)=10^{-3}$ and $\phi\left(0\right)=10^{-6}$.

It is important to remark that the study we have
undertaken here is based on an equation of state
and a thermodynamical potential ($\Omega_{e}$) that
come from a one-loop approximation.
This approximation may not be appropriate in regimes in
which the magnetic field overcomes its critical value.
This issue deserves a separate study, possibly in the context of
non-perturbative calculations at high magnetic
fields, which  would mean a different approach that may be applicable
to the magnetized gas that we have examined
here (see \cite{Ayala:2004dx} and references quoted therein).

\section{Conclusion}
We have examined the dynamical and thermodynamical behavior
of a magnetized, self--gravitating
electron gas at finite temperature, taken as the
source of a simplified Bianchi I space--time represented by a
Kasner metric, which is the simplest geometry that allows
us identify the magnetic field as the main source of
anisotropy. We regard this configuration as a toy model
that roughly approximates a grand canonical subsystem
of a magnetized electron source in the conditions prevailing
near of the center and rotation axis of a compact
object (in which spatial gradients of physical and kinematic
variables may be regarded as negligible).
The resulting Einstein--Maxwell field equations were
transformed into a system of non--linear autonomous
evolution equations, which were solved numerically
in the collapsing regime for a chemical potential:
$\tilde \mu=2m_{e}$, a magnetic field in the range
$\hbox{B}\thicksim 10^{7}-10^{8}\hbox{G}$ and temperatures
$\hbox{T}\thicksim10^{4}-10^{7}\hbox{K}$.

For all initial conditions that we considered the
gas evolves into a collapsing singularity, which can be
(depending on the initial conditions) isotropic (``point--like'') or
anisotropic (``cigar--like''). In all the studied cases the
isotropic configuration is the most stable, as the collapsing time is larger.
This result may be connected with the stability of volume
elements in less idealized
configurations, an issue that is outside the scope of this paper and deserves a proper examination elsewhere.

The collapse time decreases as the initial magnetic
field increases, but we did not find a direct
proportionality relation between this time and the initial
temperature (this may be a consequence of
having assumed a high density regime).

The behavior of the temperature as the collapse proceeds also
depends on initial conditions: for initial
temperatures $\hbox{T}\thicksim10^{4}-10^{6}\hbox{K}$ the
temperature decreases, while for values $\hbox{T}\gtrsim10^{7}\hbox{K}$ the
 temperature increases. As mentioned previously,
this difference in the temperature behavior can be explained
by the predominance of the magnetic effects (alignment of electron spins with
the magnetic field) over the effects of temperature
(random motion) during the evolution of the gas in the regime
characterized by low temperature and strong
magnetic field.

The behavior of the magnetization and energy density
was studied by means of the numerical solutions of the system.
We found a monotonic
relation between the magnetization
and the magnetic field for high temperature values,
but this relation does not occur for low temperature values.
For small temperature values, when the magnetic field
is also small, the electron gas exhibits the
Haas-van Alphen oscillations. These results, obtained in a
time-dependent system, agree with those obtained in
\cite{TAndersen,Oliver} for a time-independent system.

In all the studied cases the energy density increases
as the collapse process, even when the system
is cooled. This occurs because the particle number
density and the chemical potential increase during the collapse, causing
 the energy density to grow even when the temperature
drops.

The study we have presented can be readily applied
to examine hadronic systems (complying with suitable
balance conditions and adequate chemical potentials).
The methodology we have used can also serve as
starting point to study the origin and the dynamics
of primordial  cosmological magnetic fields. These
potential extensions of the present work are already
under  consideration for future articles.

\begin{acknowledgement}
The work of A.P.M, A.U.R and I.D has been supported by
\emph{Ministerio
de Ciencia, Tecnolog\'{\i}a y Medio Ambiente} under the grant CB0407
and the ICTP Office of External Activities through NET-35. APM
acknowledges
to Prof R. Ruffini for his hospitality and financial support at
International
Center for Relativistic Astrophysics Network-ICRANET. R.A.S. and
A.U.R. acknowledge support from the research
grant \emph{SEP--CONACYT--132132}, and the TWAS-CONACYT fellowships.
\end{acknowledgement}

\appendix
\section{Contributions to EOS} \label{EOS contribution}

The thermodynamical potential of an electron gas has two
contributions, see equation (\ref{contrib Omega}), and thus we can express the magnetization
as $M_e=M_{SQFT}+M_{QFT}$. Taking it into account we can rewrite the EOS
(\ref{eq:sistadimen-a})-(\ref{eq:vincadimen}) in the following form:
\begin{eqnarray}\label{Pz+CPz}
P_{\|}&=& -\Omega^{SQFT}-BM_{SQFT}
\underbrace{-\Omega^{QFT}-BM_{QFT}+\frac{B^{2}}{2}} \nonumber
\\
& \equiv & \overline{P}_{\|}\left(B,\mu,T\right) +\Delta
P_{\|}\left(B\right), \label{Pz+CPz}
\\
P_{\bot}&=&-\Omega_{SQFT}
\underbrace{-\Omega_{QFT}-\frac{B^{2}}{2}}\nonumber
\\
& \equiv & \overline{P}_{\bot}\left(B,\mu,T\right) +\Delta
P_{\bot}\left(B\right),\label{Px+CPx}
\\
U&=&-\Omega_{SQFT}+TS+\mu N
\underbrace{-\Omega_{QFT}+\frac{B^{2}}{2}}\nonumber
\\
& \equiv & \overline{U}\left(B,\mu,T\right) +\Delta
U\left(B\right),\label{U+CU}
\end{eqnarray}
where we identify by $\overline{P}_{\bot}$, $\overline{P}_{\|}$ and
$\overline{U}$ the terms corresponding to our approximate EOS, while
$\Delta P_{\|}$, $\Delta P_{\bot}$ and $\Delta U$ represent those
that have been neglected. Note that these last terms are
only field--dependent.

In order to analyze the contribution of $\Delta P_{\|}$, $\Delta
P_{\bot}$ and $\Delta U$ to its respective EOS, we evaluate the
expressions $\overline{P}_{\|} / \Delta P_{\|}$, $\overline{P}_{\bot} / \Delta P_{\bot}$
and $\overline{U} / \Delta U$ in the sets of solutions obtained
previously. In all the cases the numerical results
show that these terms are not significant for values
of magnetic field less than or equal to the critical
magnetic field.

\begin{figure}
\includegraphics[scale=0.35]{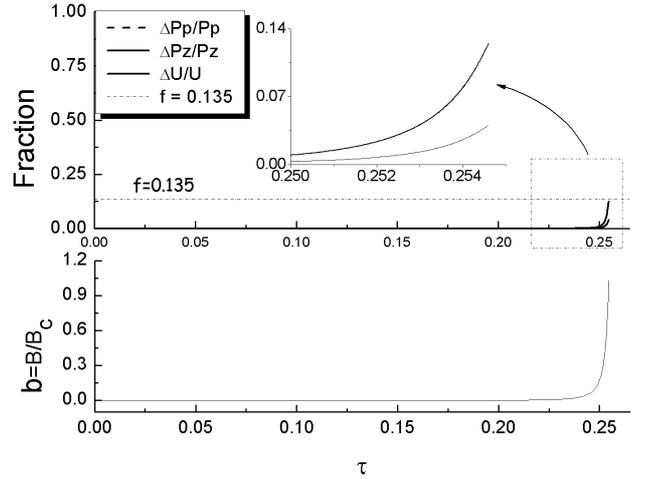}
 \caption{\label{fig CorrEOS} In the upper (lower) part are shown the fraction of contribution to EOS of the neglected terms (values of magnetic field) versus proper time.}
\end{figure}

To exemplify that mentioned above, in the figure \ref{fig CorrEOS} we show the
fractions of the EOS  neglected during the evolution of the gas
starting from
$\phi(0)=10^{-3}$, $b(0)=5\times10^{-5}$, $\mu(0)=2$,
$S_{2}(0)=0$
and $S_{3}(0)=1$. As we can observe these values remain near to zero
for $B\lesssim B_{c}$. The graph of magnetic field
was included for comparison.

\end{document}